\documentstyle[aps]{revtex}

\begin{document}

\noindent {\bf RE}\noindent \noindent \noindent \noindent \noindent
\noindent {\bf CALCULATION OF THE SPECTRUM OF THE RADIATION EMITTED\ DURING\
GRAVITATIONAL\ COLLAPSE}

\bigskip

\bigskip \noindent \noindent {\bf A\ Calogeracos}\noindent (*)

\noindent {\it Division of Theoretical Mechanics, Hellenic Air Force Academy
TG1010, Dekelia Air Force Base, Greece}

\medskip and

\noindent {\it Department of Physics, University of Sussex, Brighton, BN1 9QH%
}

\noindent \noindent May 2004

\bigskip \noindent (*) acal@hol.gr, a.calogeracos@sussex.ac.uk

quantum field theory, gravitational collapse

{\bf Abstract}

We recalculate the spectrum of radiation emitted from a collapsing star. We
consider the simple model of a spherical star consisting of pressure-free
dust and we derive the thermal spectrum via a systematic asymptotic
expansion of the complete Bogolubov amplitude. Inconsistencies in previous
derivations are pointed out.

\bigskip

\bigskip

\bigskip

More than a quarter of a century has passed since Hawking's remarkable
suggestion \cite{HAW} that a star collapsing to a black hole gives rise to
radiation emission at a steady rate characterized by the black body
spectrum. That particle creation takes place is not in itself surprising.
Let us take the case of a spherical star with radius initially larger than
its Schwarzschild radius that eventually collapses contracting to a point
(according to classical gravitation). If we consider a quantized scalar
photon field then the quantum spaces of the $in$ states (before the
initiation of collapse) and of the $out$ states (after collapse is
completed) are certainly different. Hence particle creation clearly takes
place and is determined by the Bogolubov $\alpha (\omega ,\omega ^{\prime })$
and $\beta (\omega ,\omega ^{\prime })$ amplitudes. The fact that the
spectrum is that of a black body is indeed noticeable, and ties up with the
somewhat earlier results on black hole thermodynamics; see the textbook by
Misner, Thorne and Wheeler \cite{MI} for a detailed treatment. Shortly after
Hawking's work on black hole collapse two classic papers by Fulling and
Davies \cite{FD}, \cite{DF} appeared. The authors demonstrated an
illuminating analogy, physical as well as mathematical, between
gravitational collapse and the seemingly rather different problem of a
perfect mirror starting from rest and accelerating for an infinite time; see
the textbook by Birrell and Davies \cite{BD} for a comprehensive review of
particle production by both mirrors and black holes. We shall see that the
Fulling and Davies effect provides a very good guide as to what happens in
gravitational collapse. At the root of both effects lies the fact that there
is a constant energy flux at late times as shown in both \cite{HAW} and \cite
{DF}. This does signal the presence of a thermal spectrum. It also provides
a physical interpretation of the logarithmic divergence that results upon
integration of (\ref{bb21}) over $\omega ^{\prime }$ in order to obtain the
spectrum $n(\omega )$ (the divergence will presumably be cured by a more
complete theory). From the mathematical point of view Hawking's result is
due to the special behaviour of the photon modes near the stellar surface
just before the horizon is formed and his derivation makes heavy use of the
asymptotic (i.e. near the horizon) form of the modes. The singular behaviour
of the modes in this regime has in turn given rise to various statements in
the literature that are not strictly correct. For example there have been
references to ''parts of $\alpha (\omega ,\omega ^{\prime })$ and $\beta
(\omega ,\omega ^{\prime })$ that relate to the steady-state regime at late
times'' (\cite{dew} p. 327). However the Bogolubov amplitudes are global
constructs and the distinction between early and late times does not make
strict sense. The special role of late times does of course manifest itself
in the calculation of {\it local} quantities like $\left\langle T_{\mu \nu
}\right\rangle $. The point of view that we are adopting is the one dictated
by quantum-mechanical orthodoxy, namely that particle production is a global
process (\cite{HAW} p. 216) and that once initial and final states are
specified (in the present case the initial state being the {\it in} vacuum)
then the full standard expression for $\beta (\omega ,\omega ^{\prime })$ is
able to deliver the answer. Of course at some point in the calculation of $%
\beta (\omega ,\omega ^{\prime })$ \noindent the special role of the horizon
will show up but this must be taken care of by the mathematics without any
ad hoc statements. In \cite{calo} we raised a criticism in connection to the
calculation of the black body spectrum in \cite{DF} and the purpose of this
note is to show that the points arising in \cite{calo} are also valid in the
case of gravitational collapse. As in the case of the Fulling and Davies
effect we shall perform a systematic asymptotic expansion of the amplitude
for large $\omega ^{\prime }$ and we shall show that as far as this
particular calculation is concerned we can free ourselves from distinctions
between early and late times, or transient versus steady state radiation.
For an expanded version which examines in detail the points raised in this
letter and which also to some extent duplicates results already found in the
literature see \cite{calo3}. Note: in what follows $\not{h}=G=c=1$.

Our point is best illustrated in a concrete case and for the sake of
simplicity we examine a collapsing spherical star consisting of
pressure-free dust (\cite{WEI} chapter 11, sections 8 and 9). For the
description of collapse one uses comoving coordinates ($t,r,\theta ,\varphi $%
) for the region inside the star and Schwarzschild coordinates ($\bar{t},%
\bar{r},\theta ,\varphi $) for the outside region. Since the collapse is
pressure-free each dust particle falls freely and thus the time $t$ stands
for the proper time registered by the particle in question. The two sets of
coordinates are smoothly matched at the stellar surface{\it . }The time $t$
registered by a surface particle and the radius $\bar{R}$ are given
parametrically in terms of the cycloidal variable (\cite{WEI} equations
1.9.25, 26). The Schwarzschild time $\bar{t}$ for the surface particle is
given in terms of the cycloidal variable by a more complicated expression
(see \cite{chandra} chapter 3, equation 107). It will prove crucial that we
specify the initial conditions of the collapse. Assume that the dust is held
stable until time $\bar{t}=0$ and is then allowed to contract under the
action of gravity. Let $M$ be the mass of the star, $a$ its initial radius,
and define a quantity $k$ (as in \cite{WEI}) by 
\begin{equation}
2M=ka^{3}  \label{co5}
\end{equation}

\noindent We shall consider the case of initially dilute matter $a>>M$,
which implies via (\ref{co5}) 
\begin{equation}
ka^{2}<<1  \label{co31}
\end{equation}

\noindent Hence at early times spacetime is essentially flat and the metric
is simply expressed in terms of advanced and retarded Eddington-Finkelstein
coordinates $v=\bar{t}+\bar{r},u=\bar{t}-\bar{r}$%
\begin{equation}
ds^{2}=dudv-\bar{r}^{2}d\Omega  \label{co53}
\end{equation}
(generally it is the Regge-Wheeler coordinate $\bar{r}^{*}$ that enters in
the definition of $v$ and $u$ but because of (\ref{co31}) there is little
difference between radial and Regge-Wheeler coordinates at early times). At
late times we need the metric expressed in terms of Kruskal coordinates; for
the relevant definitions see pages 20-21 of \cite{TOWN} and also chapter 31
of \cite{MI}: 
\begin{equation}
ds^{2}=\frac{32M^{3}}{\bar{r}}e^{-\bar{r}/2M}d{\it U}d{\it V}-\bar{r}%
^{2}d\Omega ,{\it U}=-Ee^{-u/4M},{\it V}=\frac{e^{v/4M}}{E},{\it UV}=-\left( 
\frac{\bar{r}-2M}{2M}\right) e^{\bar{r}/2M}  \label{co51}
\end{equation}

\noindent The presence of $E$ in (\ref{co51}) reflects an appropriate
translation in Schwarzschild spacetime $\bar{t}$ and its precise value
depends on the parameters of the collapse.

We turn to the Klein-Gordon equation satisfied by a scalar field in curved
spacetime. In this we follow section 8.1 of \cite{BD} and p. 395 of \cite
{Brout}. A scalar mode corresponding to angular momentum $l$ and satisfying
the Klein-Gordon equation is written in the form 
\begin{equation}
\psi _{l}=\left( \sqrt{4\pi }r\right) ^{-1}\phi _{l}(r)Y_{l}^{m}(\theta
,\varphi )  \label{co25}
\end{equation}

\noindent As far as conceptual purposes and technical details are concerned
it suffices to restrict ourselves to $s$ waves. Following the reasoning in 
\cite{Brout} we neglect the centrifugal barrier (present even for $s$
waves). Then the Klein-Gordon equation reduces to 
\begin{equation}
\left( \frac{\partial ^{2}}{\partial t^{2}}-\frac{\partial ^{2}}{\partial
r^{*2}}\right) \phi =0  \label{co29b}
\end{equation}

\noindent \noindent (in what follows the subscript $0$ in $\phi $ is
suppressed). Since $\psi _{0}(r=0)$ must be finite it follows from (\ref
{co25}) that 
\begin{equation}
\phi (r=0)=0  \label{co28}
\end{equation}

\noindent In terms of the Eddington-Finkelstein coordinates equation (\ref
{co29b}) takes the form of the massless scalar equation (8) of \cite{calo}.
This, together with boundary condition (\ref{co28}) on the trajectory of the
centre of the star, reduces the problem to that of a perfect mirror
following a trajectory of the type examined in \cite{calo}. In particular
the {\it in }and {\it out} modes $\varphi _{\omega }(u,v)$, $\bar{\varphi}%
_{\omega }(u,v)$ take the form of (15) and (16) of \cite{calo} respectively: 
\begin{equation}
\varphi _{\omega }(u,v)=%
{\displaystyle {i \over 2\sqrt{\pi \omega }}}%
\left( \exp (-i\omega v)-\exp \left( -i\omega p(u)\right) \right)  \label{e3}
\end{equation}

\noindent 
\begin{equation}
\bar{\varphi}_{\omega }(u,v)=%
{\displaystyle {i \over 2\sqrt{\pi \omega }}}%
\left( \exp \left( -i\omega f(v)\right) -\exp \left( -i\omega u\right)
\right)  \label{e5}
\end{equation}

\noindent Recall that apart from the modes $\bar{\varphi}_{\omega }(u,v)$ we
should include modes $\bar{q}_{\omega }(u,v)$ that contain no outgoing
component; instead they are confined inside the black hole (the letter $q$
complies with the notation in Hawking \cite{HAW}, \cite{HAW2}). The modes $%
\bar{q}_{\omega }(u,v)$ are undetectable by an outside observer, but they
are needed to make the set $\bar{\varphi}_{\omega },\bar{q}_{\omega }$
complete. Correlations between the two sets is part of the much discussed
information problem \cite{preskill}, \cite{HEL}. (Recall the analogy with
the Fulling-Davies model: the mirror trajectory in \cite{calo} is asymptotic
to the null line $v=\ln 2$ and thus does not cut all the characteristics of
the wave equation; it leaves out the ones with $v>\ln 2$. The ensuing lack
of completeness has been pointed out by Nikishov and Ritus \cite{NR}.)

To proceed with the calculation we need information on the function $u=f(v)$
appearing in the expression for $\bar{\varphi}_{\omega }(u,v)$. It will
transpire that as far as large $\omega ^{\prime }$ asymptotics is concerned
information pertaining to late and early times is sufficient. Consider an
incoming ray $l(i)$ (figure 1 of \cite{calo3}) with $v$ close to $v_{H}$
which upon reflection at the centre turns to an outgoing ray $l^{\prime }(i)$
with Kruskal coordinate ${\it U}$ close to ${\it U}=0$. For a given ${\it U}$
we determine the corresponding $u$ via the second of (\ref{co51}), and we
then connect $u$ of $l^{\prime }(i)$ to $v$ of $l(i)$ via $f$. Recall (see
p. 29 of \cite{TOWN} for proof) that ${\it U}$ is an affine parameter for a
null geodesic ${\it U}=${\it constant}. By a standard argument (p. 126 of 
\cite{TOWN}, pages 208-9 of \cite{HAW}) we obtain 
\begin{equation}
u=f(v)\simeq -\frac{1}{4M}\ln \left( \frac{v_{H}-v}{E}\right) ,v\rightarrow
v_{H}  \label{co38}
\end{equation}

We turn to the early stage of the collapse (figure 3 of \cite{calo3}).
Combining the expressions for $\bar{t}$ and $\bar{R}$ in terms of the
cycloidal parameter we obtain for $\bar{t}$ small

\begin{equation}
\bar{R}=a\left( 1-\frac{k\bar{t}^{2}}{4}\right)  \label{co20}
\end{equation}

\noindent Thus the star's surface initially contracts according to
non-relativistic kinematics with acceleration $M/a^{2}$ (in accordance with
Newton's law of gravity). Given (\ref{co31}) the initial gravitational field
is sufficiently weak so the light rays in Schwarzschild coordinates are
straight lines. Let us consider a $\bar{t}-\bar{R}$ diagram and take an
incoming light ray $q$ which has $v=0$. The reflected ray $q^{\prime }$ has $%
u=0$ and takes time $a$ to travel from the centre of the star to the
surface. During this time the gravitational field is still weak. This may be
seen by setting $t=a$ in (\ref{co20}); then (\ref{co31}) implies that by the
time the ray emerges from the star's surface $\bar{R}$ still is almost equal
to $a$. Thus at early times the function $u=f(v)$ satisfies

\begin{equation}
u=f(v)=v,initially  \label{co23}
\end{equation}

The Bogolubov amplitude $\beta (\omega ,\omega ^{\prime })$ has the standard
form (see e.g. \cite{BD}, equations (2.9) and (3.36)) 
\begin{equation}
\beta (\omega ,\omega ^{\prime })=-i\int dz\varphi _{\omega ^{\prime }}(z,0)%
{\displaystyle {\partial  \over \partial t}}%
\bar{\varphi}_{\omega }(z,0)+i\int dz\left( 
{\displaystyle {\partial  \over \partial t}}%
\varphi _{\omega ^{\prime }}(z,0)\right) \bar{\varphi}_{\omega }(z,0)
\label{e11}
\end{equation}

\noindent where we use $z$ for $r^{*}$ to ease up on notation and make
contact with \cite{calo}. The integration in (\ref{e11}) can be over any
spacelike hypersurface. Since collapse starts at $t=0$ the choice $t=0$ for
the hypersurface is convenient. The $in$ modes evaluated at $t=0$ are given
by the simple expression (\ref{e3}) (i.e. $p(u)=u)$%
\begin{equation}
\varphi _{\omega }(u,v)=%
{\displaystyle {i \over 2\sqrt{\pi \omega }}}%
\left( \exp (-i\omega z)-\exp \left( i\omega z\right) \right) \theta \left(
z\right)  \label{e99}
\end{equation}
where the presence of $\theta \left( z\right) $ emphasizes the fact that $z$
is a radial coordinate. The $\bar{\varphi}$ modes are given by (\ref{e5})
with $f$ depending on the history of the collapse. Relation (\ref{e11}) is
rearranged in the form

\noindent 
\begin{equation}
\beta (\omega ,\omega ^{\prime })=%
{\displaystyle {1 \over 4\pi \sqrt{\omega \omega ^{\prime }}}}%
\int_{0}^{v_{H}}dz\left\{ e^{i\omega ^{\prime }z}-e^{-i\omega ^{\prime
}z}\right\} \theta \left( z\right) \left\{ \omega ^{\prime }e^{-i\omega
f}-\omega f^{\prime }e^{-i\omega f}\right\} + 
{\displaystyle {\left( \omega -\omega ^{\prime }\right)  \over 4\pi \sqrt{\omega \omega ^{\prime }}}}%
\int_{0}^{\infty }dz\left\{ e^{i\omega ^{\prime }z}-e^{-i\omega ^{\prime
}z}\right\} \theta \left( z\right) e^{i\omega z}  \label{bb9}
\end{equation}

\noindent The first and second terms in the above relation will be denoted
by $\beta _{1}(\omega ,\omega ^{\prime })$ and $\beta _{2}(\omega ,\omega
^{\prime })$ respectively. Observe that the argument of $f$ runs up to $%
v_{H} $ and this is reflected in the limits of the first integral in $\beta
_{1}(\omega ,\omega ^{\prime })$. The rays with $v<0$ do not affect the
amplitude due to the presence of the $\theta $ function. Note that $\beta
_{2}(\omega ,\omega ^{\prime })$ is of kinematic origin and totally
independent of the collapse. It was pointed out in \cite{calo} that (a) this
term is unaccountably missing from \cite{DF}, (b) that the term is in fact
instrumental in obtaining the thermal spectrum in the case of an
accelerating mirror. It will be seen that the same situation holds here
(thus strengthening the analogy between Hawking and Fulling-Davies effects).
One can see that the presence of the term $\beta _{2}(\omega ,\omega
^{\prime })$ is indeed necessary by imagining that the star is held stable
by some external means and that the radius stays constant (and large). Then
trivially $f(z)=z$, the upper limit in the first integral in (\ref{bb9}) is
infinity, and the two terms identically cancel each other (as they must
since nothing is produced).

We return to the calculation of the amplitude and recall that we are chasing
the leading asymptotic $\omega ^{\prime }\rightarrow \infty $ behaviour of
the amplitude which causes the logarithmic divergence mentioned in the
Introduction. As in \cite{calo} we get

\noindent

\begin{equation}
\beta _{2}(\omega ,\omega ^{\prime })\simeq \frac{1}{2\pi i\sqrt{\omega
\omega ^{\prime }}}  \label{bb10}
\end{equation}

To calculate $\beta _{1}(\omega ,\omega ^{\prime })$ we perform an
integration by parts to obtain 
\begin{equation}
\beta _{1}(\omega ,\omega ^{\prime })=-\frac{1}{2\pi }\sqrt{\frac{\omega
^{\prime }}{\omega }}\int_{0}^{v_{H}}dze^{-i\omega f(z)-i\omega ^{\prime }z}
\label{bb23}
\end{equation}
Observe that this is exactly the same amplitude that appears in (2.10b) of
Davies and Fulling \cite{DF}. The leading order contribution will be
examined using the method in \cite{calo}. The $\omega ^{\prime }\rightarrow
\infty $ asymptotics of the integral in (\ref{bb23})\ belongs to the
standard class of problems examined, for example, in Chapter 6 of Bender and
Orszag \cite{BO}. To bring the singularity in the integral \ to zero we make
the change of variable

\begin{equation}
z=v_{H}-x  \label{e22a}
\end{equation}

\noindent and rewrite $\beta _{1}(\omega ,\omega ^{\prime })$ in the form 
\begin{equation}
\beta _{1}(\omega ,\omega ^{\prime })=-\frac{e^{-i\omega v_{H}}}{2\pi }\sqrt{%
\frac{\omega ^{\prime }}{\omega }}\int_{0}^{v_{H}}dxe^{-i\omega
g(x)}e^{i\omega ^{\prime }x}  \label{bb24}
\end{equation}

\noindent where the function \noindent \noindent $g(x)\equiv f\left(
v_{H}-z\right) $ is defined in the range $0<x<v_{H}$ and has the properties
that follow from (\ref{co38}), (\ref{co23}): 
\begin{equation}
g(x)\simeq -4M\ln \left( \frac{x}{E}\right) ,x\rightarrow 0  \label{co42}
\end{equation}
\begin{equation}
g(v_{H})=f(0)  \label{co43}
\end{equation}
We momentarily drop prefactors and isolate the integral

\begin{equation}
I\equiv \int_{0}^{v_{H}}dxe^{-i\omega g(x)}e^{i\omega ^{\prime }x}
\label{e24b}
\end{equation}

\noindent

To obtain the asymptotic behaviour of (\ref{e24b}) for $\omega ^{\prime }$
large we use the same technique as in \cite{calo} (the two integrals being
essentially identical). We deform the integration path to a contour in the
complex plane; see Bender and Orszag {\it op cit}; chapter 6 of \cite{Abl};
also \cite{MF}, p. 610 where a very similar contour is used in the study of
the asymptotic expansion of the confluent hypergeometric. The deformed
contour runs from 0 up the imaginary axis till $iT$ (we eventually take $%
T\rightarrow \infty $), then parallel to the real axis from $iT$ to $%
iT+v_{H} $, and then down again parallel to the imaginary axis from $%
iT+v_{H} $ to $v_{H}$. The contribution of the segment parallel to the real
axis vanishes exponentially in the limit $T\rightarrow \infty $. We thus get 
\begin{equation}
I=i\int_{0}^{\infty }dse^{-\omega ^{\prime }s}e^{-i\omega
g(is)}-i\int_{0}^{\infty }dse^{i\omega ^{\prime }\left( v_{H}+is\right)
}e^{-i\omega g(v_{H}+is)}  \label{co44}
\end{equation}
In both integrations the dominant contribution comes from the region where $%
s\simeq 0$. Using (\ref{co43}) the second integral (including the minus sign
in front) takes the form 
\begin{equation}
-ie^{i\omega ^{\prime }v_{H}}e^{-i\omega f(0)}\int_{0}^{\infty }dse^{-\omega
^{\prime }s}=-i\frac{e^{i\omega ^{\prime }v_{H}}e^{-i\omega f(0)}}{\omega
^{\prime }}  \label{co45}
\end{equation}

\noindent In the first integral we use the asymptotic form (\ref{co42})
valid for small $x$ and write 
\[
\exp \left( -i\omega g\left( is\right) \right) =\exp \left( i4M\omega \ln
\left( \frac{is}{E}\right) \right) =\left( \frac{is}{E}\right) ^{i4M\omega
}=\exp \left( -\frac{\pi }{2}4M\omega \right) \left( \frac{s}{E}\right)
^{i4M\omega } 
\]

\noindent where we took the branch cut of the function $x^{i\omega }$ to run
from zero along the negative $x$ axis, wrote $x^{i\omega }=\exp \left(
i\omega \left( \ln x+i2N\pi \right) \right) $ and chose the branch $N=0$.
Thus 
\begin{equation}
i\int_{0}^{\infty }dse^{-\omega ^{\prime }s}e^{-i\omega g(is)}=i\exp \left( -%
\frac{\pi }{2}4M\omega \right) E^{-i4M\omega }\int_{0}^{\infty }dse^{-\omega
^{\prime }s}\left( s\right) ^{i4M\omega }=i\exp \left( -\frac{\pi }{2}%
4M\omega \right) E^{-i4M\omega }\frac{\Gamma \left( 1+i4M\omega \right) }{%
\left( \omega ^{\prime }\right) ^{1+i4M\omega }}  \label{co46}
\end{equation}

\noindent We substitute (\ref{co45}) and (\ref{co46}) in (\ref{co44}) and
then in (\ref{bb24}) to get

\begin{equation}
\beta _{1}(\omega ,\omega ^{\prime })=-i\frac{e^{-i\omega v_{H}}}{2\pi \sqrt{%
\omega \omega ^{\prime }}}\exp \left( -\frac{\pi }{2}4M\omega \right)
E^{-i4M\omega }\frac{\Gamma \left( 1+i4M\omega \right) }{\left( \omega
^{\prime }\right) ^{i4M\omega }}-\frac{e^{-i\omega f(0)}}{i2\pi \sqrt{\omega
\omega ^{\prime }}}  \label{co47}
\end{equation}

\noindent Collecting (\ref{bb10}) and (\ref{co47}) we get 
\begin{equation}
\beta \left( \omega ,\omega ^{\prime }\right) =-i\frac{e^{-i\omega v_{H}}}{%
2\pi \sqrt{\omega \omega ^{\prime }}}\exp \left( -\frac{\pi }{2}4M\omega
\right) E^{-i4M\omega }\frac{\Gamma \left( 1+i4M\omega \right) }{\left(
\omega ^{\prime }\right) ^{i4M\omega }}-\frac{e^{-i\omega f(0)}}{i2\pi \sqrt{%
\omega \omega ^{\prime }}}+\frac{1}{2\pi i\sqrt{\omega \omega ^{\prime }}}
\label{co48}
\end{equation}

The day is saved by (\ref{co23}) which causes an exact cancellation of the
last two terms. The first term on its own immediately leads to the black
body spectrum. Taking its modulus, squaring, and using the property 
\[
\left| \Gamma \left( 1+iy\right) \right| ^{2}=\pi y/\sinh \left( \pi
y\right) 
\]

\noindent we get 
\begin{equation}
\left| \beta \left( \omega ,\omega ^{\prime }\right) \right| ^{2}=\frac{4M}{%
2\pi \omega ^{\prime }}\frac{1}{e^{8\pi \omega M}-1}  \label{bb21}
\end{equation}

It is instructive to compare with earlier work on the subject. One often
starts (see e.g. \cite{BD} p. 108, \cite{DF} equation (2.10b)) with
expression (\ref{bb23}) thus unjustifiably disregarding $\beta _{2}(\omega
,\omega ^{\prime })$ (in this respect cf the remarks following (\ref{bb9})
above). One then uses (\ref{co42}) (or its equivalent (\ref{co38}))
throughout the range of integration and not just asymptotically near the
horizon where it is valid. One thus gets dropping constant prefactors 
\begin{equation}
\beta _{1}(\omega ,\omega ^{\prime })\approx \int_{0}^{v_{H}}dx\left( \frac{x%
}{E}\right) ^{i\omega 4M}e^{i\omega ^{\prime }x}  \label{a6}
\end{equation}
That (\ref{a6}) is an inadequate approximation to the original integral may
easily be seen by the fact that the use of (\ref{co42}) has changed the
behaviour of the integrand at $x=v_{H}$ (the non-singular end). From a
physical point of view the use of (\ref{co38}) at $z=0$ induces a
discontinuity in the {\it derivative} $f^{\prime }(v)$ at $t=0$
(corresponding in the Fulling-Davies model to an infinite acceleration of
the mirror). One then proceeds to rewrite (\ref{a6}) by rescaling $\omega
^{\prime }x\rightarrow x$%
\begin{equation}
\beta _{1}(\omega ,\omega ^{\prime })\approx E^{-i\omega 4M}\left( \omega
^{\prime }\right) ^{-i\omega 4M-1}\int_{0}^{\omega ^{\prime
}v_{H}}dxx^{i\omega 4M}e^{ix}  \label{680a}
\end{equation}

\noindent Since one is chasing the ultraviolet divergence one simply sets $%
\omega ^{\prime }v_{H}=\infty $, changes variable $\rho =i\sigma $ and
rotates in the complex plane to get (\ref{680a}) in the form 
\begin{equation}
\beta _{1}(\omega ,\omega ^{\prime })\approx E^{-i\omega 4M}\left( \omega
^{\prime }\right) ^{-i\omega 4M-1}e^{-\frac{\pi }{2}\omega
4M}\int_{0}^{\infty }d\sigma e^{-\sigma }\sigma ^{i\omega }  \label{e18}
\end{equation}

\noindent This reasoning is implicit in Hawking's statement (\cite{HAW}, p.
209, lines preceding (2.19)) on the Fourier transform of the amplitude. Note
that setting $\omega ^{\prime }v_{H}=\infty $ certainly does not{\it \ }%
amount to a systematic expansion in $\left( \omega ^{\prime }\right) ^{-1}$.
The $\sigma $ integration yields $\Gamma (1+i\omega )$ and one thus obtains
the form for the $\beta _{1}$ amplitude leading to the black body spectrum.
On the other hand the step from (\ref{680a}) to (\ref{e18}) is again
questionable. Integral (\ref{a6}) can be performed exactly in terms of the
confluent hypergeometric function and the asymptotic estimate for large $%
\omega ^{\prime }$ may be examined afterwards. Indeed let us rescale the
variable in (\ref{a6}) $x\rightarrow x/v_{H}$ and rewrite 
\begin{eqnarray}
\beta _{1}(\omega ,\omega ^{\prime }) &\approx &v_{H}^{i\omega
4M+1}E^{-i\omega 4M}\int_{0}^{1}dxe^{i\omega ^{\prime }v_{H}x}x^{i\omega 4M}=
\label{bb18} \\
&=&v_{H}^{i\omega 4M+1}E^{-i\omega 4M}\frac{1}{i\omega +1}M\left( 1+i\omega
4M,2+i\omega 4M,i\omega ^{\prime }v_{H}\right)  \nonumber
\end{eqnarray}
where $M$ is the confluent hypergeometric. We can now examine the asymptotic
limit of (\ref{bb18}) for large $\omega ^{\prime }$. The asymptotic limit of
the confluent $M(a,b,i\left| z\right| )$ for large values of $\left|
z\right| $ is given by item 13.5.1 of \cite{AS} ($z\equiv i\omega ^{\prime
}v_{H}$). In the case $b=a+1$ some simplifications occur and we get 
\begin{equation}
M\left( 1+i\omega ,2+i\omega ,i\left| z\right| \right) \simeq -\left(
1+i\omega \right) e^{i\left| z\right| }\frac{i}{\left| z\right| }+i\Gamma
\left( 2+i\omega \right) \frac{e^{-\frac{\pi \omega }{2}}}{\left| z\right|
^{1+i\omega }}  \label{bb20}
\end{equation}
(other terms are down by higher powers of $1/\left| z\right| $). The second
term of the above relation combined with the prefactors in (\ref{bb18}) does
feature the $\Gamma \left( 1+i\omega \right) e^{-\frac{\pi \omega }{2}}$
factor characteristic of the black body spectrum. The reason for the
discrepancy between (\ref{e18}) and (\ref{bb20}) lies in the fact that one
should first evaluate the integral in terms of the confluent and then take
the $\omega ^{\prime }\rightarrow \infty $ limit rather than take the limit
first. The rotation in the complex plane stumbles upon the Stokes phenomenon
for the confluent (different limits for $\left| z\right| \rightarrow \infty $
depending on $\arg z)$). In short the black body spectrum (\ref{e18}) is
obtained via steps identical to the ones in \cite{DF} and the criticism
raised in \cite{calo} with respect to the latter apply to the case of
gravitational collapse too. In this respect note the remark in \cite{DF}
(the statement in parentheses following (2.10)) where doubts are implicit as
to the validity of the aforementioned mathematical steps. One possible
reason for the confusion may arise from a {\it mathematical }analogy with
the Unruh effect (see e.g. chapter 5 of the book by Wald \cite{Wald} for an
excellent review). Equation (\ref{680a}) above is essentially identical to
Wald's (5.1.11). Note however that Wald's (5.1.5) is {\it exact }for {\it %
all }$V$ because of the kinematics of the problem and that the upper limit
in (5.1.11) is strictly infinity because this is how the coordinates are
defined. On the other hand the time $t=0$ (when collapse starts or when the
Fulling-Davies mirror starts moving) and which is essentially responsible
for the appearance of $\beta _{2}(\omega ,\omega ^{\prime })$ (\ref{bb10})
simply does not appear in the Unruh problem (where the observer is forever
accelerated).

\noindent \noindent {\bf Acknowledgment}

The author is greatly indebted to Professor S\ A\ Fulling for \noindent his
help.

\end{document}